# Interface-Controlled Ferroelectricity at the Nanoscale

Chun-Gang Duan,[1,2,3] Renat F. Sabirianov,[2,3] Wai-Ning Mei,[2,3] Sitaram S. Jaswal,[1,3] and Evgeny Y. Tsymbal[1,3*]

[1] *Department of Physics and Astronomy, University of Nebraska, Lincoln, Nebraska 68588*
[2] *Department of Physics, University of Nebraska, Omaha, Nebraska 68182*
[3] *Center for Materials Research and Analysis, University of Nebraska, Lincoln, Nebraska 68588*

Recent experimental results demonstrate that in thin films ferroelectricity persists down to film thickness of a few unit cells. This finding opens an avenue for novel electronic devices based on ultrathin ferroelectrics, but also raises questions about factors controlling ferroelectricity and the nature of the ferroelectric state at the nanoscale. Here we report a first-principles study of $KNbO_3$ ferroelectric thin films placed between two metal electrodes, either $SrRuO_3$ or Pt. We show that the bonding at the ferroelectric-metal interface imposes severe constraints on the displacement of atoms, destroying the bulk tetragonal soft mode in thin ferroelectric films. This does not, however, quench local polarization. If the interface bonding is sufficiently strong the ground state represents a ferroelectric double-domain structure, driven by the intrinsic oppositely-oriented dipole moments at the two interfaces. Although the critical thickness for the net polarization of the $KNbO_3$ film is finite – about 1 nm for Pt and 1.8 nm for $SrRuO_3$ electrodes – local polarization persists down to the thickness of a unit cell.

Ferroelectric materials are very promising for various technological applications such as dynamic random access memories and non-volatile binary data storage media.[1,2] A continuing demand to further miniaturize electronic devices brings up a problem of the existence of ferroelectricity at the nanometre scale.[3,4,5] Recent experimental results demonstrate that in thin films ferroelectricity persists down to film thickness of a few unit cells,[6,7,8,9,10] which opens a new direction for novel electronic devices such as ferroelectric tunnel junctions.[11,12]

In bulk ferroelectric perovskites, the spontaneous uniform electric polarization arises due to the displacements of the negatively charged oxygen ions relative to the positively charged ions, breaking the cubic symmetry and producing a net electric dipole moment. These displacements represent a zone-centre vibrational mode with a vanishing frequency and are therefore called a "soft mode".[13] Although the ferroelectric soft-mode is a collective motion of atoms in bulk material, there are theoretical evidences that the length scale for ferroelectricity can be reduced down to a nanometre range. For example, *ab-initio* calculations carried out for $KNbO_3$ predict that the high-symmetry cubic phase is unstable with respect to the displacement of a 2 nm chain of atoms along the [001] direction, producing local polarization.[14] Recent first-principles theoretical studies show that the critical size for ferroelectricity in perovskite films can be as small as a few lattice parameters.[15,16,17,18]

The existence of the critical thickness for ferroelectricity in perovskite thin films is usually explained by depolarizing fields produced by polarization charges accumulated on the two surfaces of the film.[19] If the ferroelectric film is placed between two metal electrodes the polarization charges are screened. However, when the film thickness is sufficiently reduced the screening becomes incomplete and at some critical thickness the electrostatic energy associated with the depolarizing fields overcomes the energy gained due to ferroelectric ordering. At this thickness the ferroelectric state becomes unstable. Recent first principles calculations of the soft-mode ferroelectric instability in $BaTiO_3$ films sandwiched between two $SrRuO_3$ electrodes support this mechanism for critical thickness.[17]

The above arguments, though important, do not, however, take into account the effect of interfaces which significantly influence soft mode displacements of the interfacial atoms in the ferroelectric due to bonding of these atoms to adjacent atoms in the metal electrodes. If the interface bonding is sufficiently strong the presence of interfaces imposes restrictions on the soft-mode motion since the atoms at the boundary of the ferroelectric are pinned to the electrodes. This inevitably affects the displacements of other atoms in the ferroelectric and may completely destroy the soft mode instability.

Furthermore, the presence of interfaces reduces symmetry of a solid due to dissimilar local environment at the interfaces compared to the bulk. The reduced symmetry may produce electric dipoles at the interfaces even for a paraelectric state. The interface dipoles can be created as a result of interface strain, atomic rippling, non-stoichiometry, and/or modified valence at the interface. In addition, a charge transfer between metal and insulator (semiconductor) due to different work functions can produce an electric dipole. For a dielectric film placed between same two metals with identical termination at the interfaces, these dipole moments by symmetry are normal to the interfaces and are pointed in the opposite directions.

If the magnitude of the interface dipole moments is comparable to the respective dipole moments in the bulk ferroelectric and if the interface bonding is strong enough, the oppositely-oriented dipoles at the two interfaces may favour a ferroelectric *double-domain structure*, as is shown schematically in Fig.1. This happens when the energy required for destroying the intrinsic dipoles at the interfaces exceeds the energy needed to create a domain wall between the two ferroelectric domains with opposite directions of the polarization. In the latter case the double domain is the minimum energy structure which persists at any film thickness. We note that this structure is different from the stripe domain structure occurring in ultrathin ferroelectric films to minimize the energy of depolarizing fields.[8]

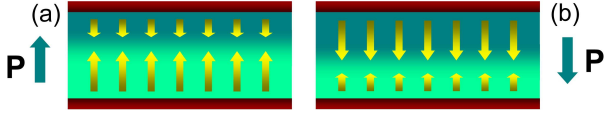

**Fig. 1** Schematic representation of a double-domain structure in a ferroelectric film placed between metal electrodes. The double domain structure is produced due to intrinsic oppositely-oriented interfacial dipoles and strong interface bonding and has the domain wall parallel to the interfaces. **a**, Net polarization pointing up; **b**, net polarization pointing down.

In order to elucidate the effect of interfaces on ferroelectricity in perovskite thin films, we have carried out first-principles calculations of the electronic and atomic structure of $KNbO_3$ thin films placed between two metal electrodes, either $SrRuO_3$ or Pt. The motivation for this choice of materials is as follows. Potassium niobate, $KNbO_3$, is a well-known perovskite ferroelectric oxide which has the spontaneous polarization $P_s \approx 0.37\,C/m^2$ and the Curie temperature $T_C \approx 704\,K$ at which the cubic–tetragonal transition occurs.[20] Strontium ruthenate, $SrRuO_3$, and platinum, Pt, are metals that are commonly used in ferroelectric devices as metal electrodes.[21] Bulk $KNbO_3$ has a very good lattice match with both bulk $SrRuO_3$ and bulk Pt (with a mismatch less than 2%) which allows layer-by-layer epitaxial growth with no misfit dislocations. From our perspective, the most important property is the difference in the electronic and atomic structure between $SrRuO_3$ and Pt. Bulk $SrRuO_3$ has a perovskite structure similar to $KNbO_3$ with a heteropolar covalent bonding between atoms. On the other hand, Pt is an elemental fcc metal with a dominant metallic bonding, making it very different from $KNbO_3$. We, therefore, expect very dissimilar properties of the $SrRuO_3/KNbO_3$ and $Pt/KNbO_3$ interfaces which should inevitably affect ferroelectric properties of the $KNbO_3$ film.

In our calculations we use a supercell geometry in which $SrRuO_3$ (001) or Pt (001) layers serve as metal electrodes separated by ferroelectric $KNbO_3$ (001) layers. Periodic boundary conditions of the supercell geometry impose the short-circuit condition between the electrodes. We assume that the $KNbO_3$ layers have $NbO_2$ terminations at both interfaces. In this case the $Pt/KNbO_3$ interface has the most stable structure if interfacial O atoms occupy atop sites on Pt. The $SrRuO_3/KNbO_3$ interface is most stable when interfacial Nb atoms are placed above O atoms lying within the SrO terminated interface of $SrRuO_3$. Thus, the supercells we use are $[SrO\text{-}(RuO_2\text{-}SrO)_4/NbO_2\text{-}(KO\text{-}NbO_2)_m]$ and $[Pt_{11}/NbO_2\text{-}(KO\text{-}NbO_2)_m]$, where $m$ = 2, 4, 6, 8, 10, 16. Figs. 2a,b show the atomic structures of these supercells for $m = 6$.

The total energy calculations are performed within density-functional theory using the projector augmented wave (PAW) method implemented in the Vienna *Ab-Initio* Simulation Package (VASP).[22] In the calculations the exchange-correlation potential is treated in the local density approximation using Ceperly-Alder scheme. We use the energy cut-off of 500 eV for the plane wave expansion of the PAWs and a 4x4x1 Monkhorst Pack grid for $k$-point sampling. Following the procedure of Ref.

17, we include implicitly the $SrTiO_3$ substrate by constraining the in-plane lattice constants of the supercell structures to the bulk lattice constant of $SrTiO_3$, 3.905 Å. This constraint is also applied to relax the bulk structure of $SrRuO_3$, Pt and $KNbO_3$. The obtained structures of tetragonal symmetry are then used to build up the metal/ferroelectric supercells. The thickness of the electrode layer (either $SrTiO_3$ or Pt) is fixed to five unit cells, i.e. about 20 Å. Atomic relaxations are performed until the Hellman-Feynman forces on atoms have become less than 20 meV/Å.

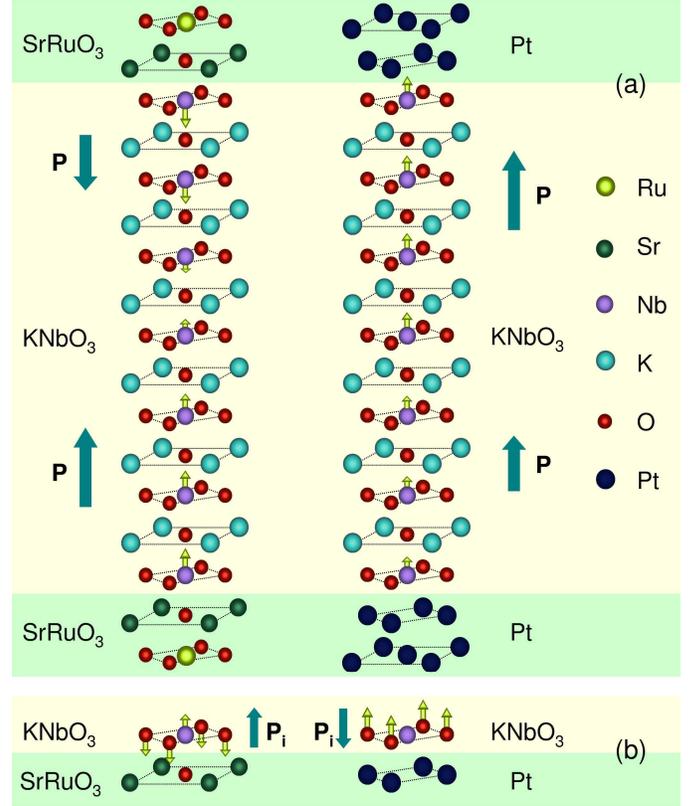

**Fig. 2** Atomic structures of $SrRuO_3/KNbO_3$ (left) and $Pt/KNbO_3$ (right) superlattices for $m = 6$. (a) Nb displacements (indicated by arrows) with respect to their bulk positions (displacements of other atoms are not shown). Only two monolayers of metal electrodes (either Pt or $SrRuO_3$) adjacent to the interfaces are shown. Double domain structure for $SrRuO_3/KNbO_3$ and a monodomain structure for $Pt/KNbO_3$ are seen from the displacements of Nb atoms as indicated by block arrows showing the direction and the magnitude of local polarization. (b) Atomic displacements at the interfaces in a paraelectric state. The intrinsic interfaces dipole moments are indicated by block arrows.

First, we analyze the electronic and atomic structure of the $SrRuO_3/KNbO_3$ and $Pt/KNbO_3$ interfaces assuming that $KNbO_3$ is in a paraelectric state. For this purpose we imposed a mirror plane on the central $NbO_2$ layer and minimized the total energy of the whole system by relaxing the distance between the metal and $KNbO_3$ layers. Then, we relaxed all the atoms at the interface, keeping rest of the atoms at fixed positions. We found that different electrodes affect differently the displacements of the interfacial atoms. For the $SrRuO_3$ electrode, the Nb atoms at the interface are pushed about 0.03 Å towards the $KNbO_3$ layer, whereas the O atoms in the same monolayer were pushed away by about 0.06 Å, as is illustrated in Fig. 2b. In contrast, for the Pt electrode, the Nb atoms tend to stay in their original positions, while the O atoms are pushed towards the KNbO$_3$ layer by about 0.10 Å (see Fig. 2b). This implies that bonding between the metal and ferroelectric atoms at the interface induces an interface dipole moment or the interface polarization, $\mathbf{P}_i$, which is electrode dependent and is oriented in the opposite directions at the two interfaces.

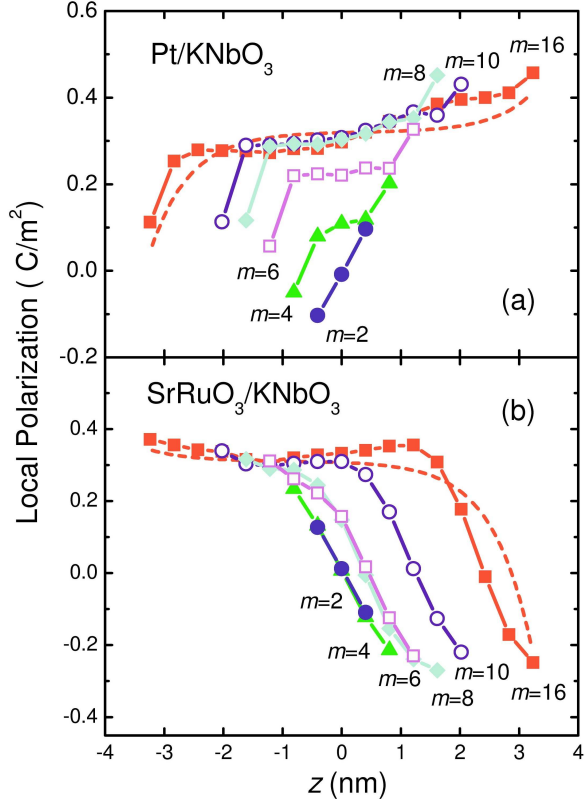

**Fig. 3** Local polarization in a ferroelectric KNbO$_3$ layer. (a) Pt electrodes; (b) SrRuO$_3$ electrodes. Positive values correspond to the polarization pointing up in Fig. 2. The midpoint of the ferroelectric layer lies at $z = 0$. Solid lines with symbols are obtained from first principles calculations for $m = 2$ (solid circles), $m = 4$ (triangles), $m = 6$ (open squares), $m = 8$ (diamonds), $m = 10$ (open circles) and $m = 16$ (solid squares). Dashed curves are results of the phenomenological theory (only curves for $m = 16$ are displayed).

We calculated the magnitude of the interface polarization using the bulk Born effective charges averaged over the KNbO$_3$ unit cell [23] and found that $P_i = 0.30$ C/m$^2$ for the SrRuO$_3$/KNbO$_3$ interface and $P_i = -0.19$ C/m$^2$ for the Pt/KNbO$_3$ interface. Here the plus sign implies that the $\mathbf{P}_i$ is pointed towards the KNbO$_3$ layer (i.e. up at the bottom interface), whereas the minus sign implies that the $\mathbf{P}_i$ is pointed away from the KNbO$_3$ layer (i.e. down at the bottom interface), as is shown in Fig. 2b. These interface polarizations are intrinsic to the interfaces and impose boundary conditions for ferroelectric instability.

We estimated the difference in the bonding strengths at the SrRuO$_3$/KNbO$_3$ and Pt/KNbO$_3$ interfaces by calculating the work of separation, $W_s$, using the method described in Ref. 24. We found that $W_s = 3.4$ J/m$^2$ for the SrRuO$_3$/KNbO$_3$ interface and $W_s = 2.1$ J/m$^2$ for the Pt/KNbO$_3$ interface. Assuming that Nb-O bonds for SrRuO$_3$/KNbO$_3$ and Pt-O bonds for Pt/KNbO$_3$ provide the dominant contribution to the work of separation, we obtained the Nb-O bond energy of about 3.2 eV and the Pt-O bond energy of about 1eV. Therefore, the Nb-O bonds in the SrRuO$_3$/KNbO$_3$ structure are stronger by more than a factor of three than the Pt-O bonds in the Pt/KNbO$_3$ structure.

To reveal the influence of these interface properties on ferroelectricity, we relaxed the constraint of reflection symmetry and minimized the total energy with respect to atomic coordinates of all the atoms in the ferroelectric. Then, using the Born charges[23] we calculated local polarization in the KNbO$_3$ film. The results are presented in Fig. 3a for Pt electrodes and in Fig. 3b for SrRuO$_3$ electrodes.

As is evident from Fig. 3a, for Pt electrodes the polarization ($P$) varies appreciably across the KNbO$_3$ layer. For example, for $m = 16$ it changes from about 0.1 C/m$^2$ at the bottom interface ($z < 0$) to about 0.43 C/m$^2$ at the top interface ($z > 0$) compared to the value of 0.33 C/m$^2$ we calculated for bulk KNbO$_3$ using Berry's phase method.[25] This strongly inhomogeneous distribution is the consequence of the asymmetry produced by the two interfaces which have the intrinsic dipole moments pointing in the opposite directions. Due to the relatively weak bonding at the Pt/KNbO$_3$ interface, the interface O and Nb atoms are not strongly coupled to Pt and are displaced in response to ferroelectric instability. Hence, the interface polarization is enhanced compared to the bulk value at the interface where the bulk polarization is pointing in the same direction as the interface dipole moment (the top interface) and is reduced at the interface where the bulk polarization is pointing in the opposite direction to the interface dipole moment (the bottom interface). As is seen from Fig.3a, with decreasing thickness of the KNbO$_3$ layer the picture does not change much down to $m = 8$. At $m = 6$ the net polarization starts to decrease as a result of the increasing relative weight of the interface energy in the total free energy of the ferroelectric. At this point the interface polarization reduces and at smaller thicknesses ($m = 4$ and $m = 2$) it changes sign at the bottom interface reflecting the orientation of the intrinsic dipole moment. The net polarization vanishes at the critical thickness of about 1 nm ($m = 2$). Note that at this thickness the local polarization represents the two oppositely-oriented dipoles at the two interfaces and a zero local polarization in the middle (Fig. 3a).

For SrRuO$_3$ electrodes, the polarization is also very inhomogeneous across the KNbO$_3$ layer (Fig. 3b). However, the striking difference in this case is that $P$ has *opposite* sign at the top and bottom interfaces for all thicknesses. This implies that the local polarization is pointing in the opposite directions at the two interfaces thereby creating a ferroelectric double-domain structure. The origin of this behaviour is relatively large oppositely-oriented intrinsic dipole moments at the two interfaces and strong interface bonding. The latter pins interfacial Nb and O atoms to their positions in the paraelectric state. Hence, the ferroelectric polarization is built according to these boundary conditions. As is seen from Fig. 3b, for $m = 16$ the double domain structure represents a relatively large domain with the polarization pointing up (positive values of $P$ in Fig. 3b) and a small domain with polarization pointing down (negative values of *P* in Fig. 3b) separated by a domain wall of 1 nm thickness. With decreasing film thickness the large domain shrinks reducing the net polarization of the film until the film thickness becomes comparable to the domain wall width. This occurs at the critical thickness of about 1.8 nm (*m* = 4) at which the net polarization vanishes. It is interesting that the local polarization is non-zero even in the absence of the net polarization and represents the double domain structure (see curves for *m* = 2 and *m* = 4 in Fig. 3b).

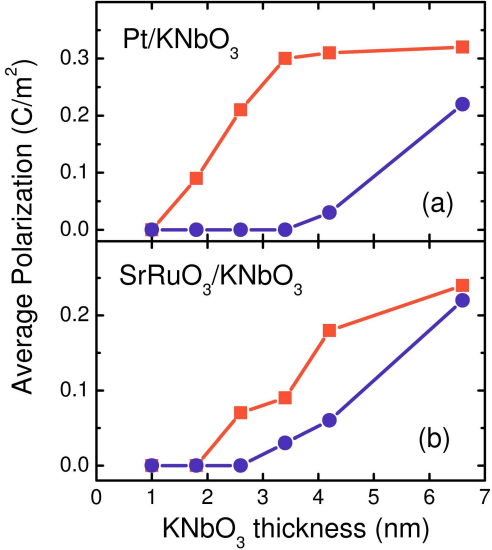

**Fig. 4** Average spontaneous polarization as a function of $KNbO_3$ film thickness. (a) Pt electrodes; (b) $SrRuO_3$ electrodes. Circles represent results of the calculation based on bulk soft-mode displacements; squares are results obtained from full relaxation of superlattices using local polarizations shown in Figs. 3a,b.

The strongly inhomogeneous displacements characterizing ferroelectric instability in ultrathin $KNbO_3$ films are very different from the soft-mode displacements typical for bulk $KNbO_3$.[14] In order to elucidate the effect of this difference on the critical thickness for ferroelectricity we carried out additional calculations in which we searched for soft-mode ferroelectric instability. For this purpose, we used the atomic structures obtained for paraelectric $KNbO_3$ and moved $KNbO_3$ atoms according to the bulk soft-mode displacements to find the energy minimum.[17] The magnitude of the polarization versus film thickness obtained from the soft-mode displacements is shown in Figs. 4a,b in comparison to the average polarization calculated from fully relaxed atomic structures. It is seen from Figs. 4a,b that above the critical thickness the magnitude of the polarization corresponding to the soft mode displacements is significantly smaller than that obtained for the fully-relaxed structures. For example, in case of Pt electrodes even at 6.4 nm thickness (m=16), the soft-mode polarization does not approach the polarization of the fully relaxed structure (Fig. 4a). Moreover, as is seen from Figs. 4a,b, the critical thicknesses of $KNbO_3$ for both Pt and $SrRuO_3$ electrodes calculated using the soft-mode description exceeds the critical thicknesses obtained from the fully-relaxed structures. For example, in case of Pt electrodes, we found the critical thickness of about 3.3 nm for soft-mode displacements versus 1 nm for the fully-relaxed structure. Thus, we conclude that the soft mode is improper description of ferroelectricity at the nanoscale. This statement is consistent with the prediction of Fridkin.[26]

Finally, we link the results of the density-functional calculations to the phenomenological theory of a ferroelectric film.[27] For this purpose we included explicitly the term which depends on the interface polarization $P_i$ in the Ginzburg-Landau-Devonshire free energy per unit area:[28]

$$F = \int_{-L/2}^{L/2} \left\{ \frac{A}{2\varepsilon_0} P^2 + \frac{B}{4\varepsilon_0^2} P^4 + \frac{D}{2}\left(\frac{dP}{dz}\right)^2 + \frac{P^2 - \bar{P}^2}{2\varepsilon_0} \right\} dz + \frac{D}{2\delta}\left\{ [P(-L/2) + P_i]^2 + [P(L/2) - P_i]^2 \right\}. \quad (1)$$

Here *L* is film thickness and $\bar{P} \equiv \frac{1}{L}\int_{-L/2}^{L/2} P(z)dz$ is the average polarization. Parameters *A* and *B* describe the bulk double-well potential profile, parameter *D* reflects the spatial variation of the polarization, and the extrapolation length $\delta$ is a phenomenological parameter associated with the derivatives of $P(z)$ at the interfaces.[27] Due to the presence of the interface dipole moments the interface contribution to the free energy contains a linear term with respect to the polarization. Variation of the free energy functional over the polarization yields the Euler–Lagrange equation

$$\varepsilon_0 D \frac{d^2 P}{dz^2} = AP + \frac{B}{\varepsilon_0} P^3 + (P - \bar{P}) \quad (2)$$

and boundary conditions

$$\left(P - \delta \frac{dP}{dz}\right)_{z=-L/2} = -P_i, \quad \left(P + \delta \frac{dP}{dz}\right)_{z=L/2} = P_i. \quad (3)$$

We solve equation (2) numerically subject to boundary conditions (3) using parameters $A = -1.42 \times 10^{-2}$ and $B = 1.14 \times 10^{-12}$ m$^3$/J which were obtained by fitting the energy-versus-polarization curve for the bulk tetragonal $KNbO_3$ structure. Parameter *D* was assumed to be the same for the two electrodes and was fixed at $D = 5.0 \times 10^{-8}$ F$^{-1}$m$^3$. $P_i$ and $\delta$ were considered as fitting parameters. From fitting the polarization distribution curves shown in Figs. 3a,b we obtained $\delta = 2$ Å and $P_i = 0.35$ C/m$^2$ for the $SrRuO_3/KNbO_3$ structure and $\delta = 20$ Å and $P_i = -0.6$ C/m$^2$ for the $Pt/KNbO_3$ structure. The resulting fitting curves are shown in Figs. 3a,b by dashed lines for the $KNbO_3$ film of 6.6 nm thickness (*m* = 16). As is seen, they are in good agreement with the first-principles results, reflecting both the asymmetry and inhomogeneity in the polarization distribution and the appearance of the double domain structure in the case of $SrRuO_3$ electrodes. The phenomenological model shows that the extrapolation length $\delta$ is related to the strength of bonding at the interface: larger $\delta$ implies weaker interface bonding. For example, we found that for $L = 6.4$ nm (*m* = 16) increasing $\delta$ from 2 Å to 8 Å eliminates the double domain structure, the requirement for the existence of which is the strong interface bonding.

The results obtained from the fitting show that in case of SrRuO$_3$ electrodes the interface polarization $P_i = 0.35$ C/m$^2$ is very close to the intrinsic interface polarization of 0.3 C/m$^2$ calculated for paraelectric KNbO$_3$, as was described above. This fact indicates that the intrinsic interface dipole moment, resulting from ionic displacements at the interface, largely controls the polarization profile in the ferroelectric layer. In case of Pt electrodes, however, the obtained interface polarization $P_i = -0.6$ C/m$^2$ is greater than the intrinsic interface polarization of $-0.19$ C/m$^2$ calculated for paraelectric KNbO$_3$. The difference between the two is large so that using $P_i = -0.19$ C/m$^2$ in the phenomenological model does not allow us to reproduce, even qualitatively, the first-principles results. The origin of this inconsistency is the electronic contribution to the intrinsic interface polarization, resulting from the charge transfer between Pt and KNbO$_3$, which was not taken into account by the Born charges. We estimated the electronic contribution to the interface dipole moments from the charge transfer between Pt and NbO$_2$ atomic monolayers adjacent to the interface. This gives a dipole moment of about $-0.2$ C/m$^2$ pointing in the same direction as the ionic dipole moment. This explains qualitatively the result of the phenomenological model. For SrRuO$_3$/KNbO$_3$ structure we found that the electronic dipole moment is negligible compared to the ionic dipole moment.

In conclusion, we have shown that the strength of bonding and intrinsic dipole moments at the interfaces control ferroelectricity in perovskite films of a nanometre thickness. The polarization profile is strongly inhomogeneous across film thickness and represents a double-domain structure if the interface bonding is sufficiently strong and if the intrinsic interface dipole moments are comparable to the bulk polarization. The critical thickness is strongly affected by the interfaces and is controlled by the domain wall width in ferroelectric films with the double-domain structure. We hope that our results will further stimulate experimental studies of ferroelectricity at the nanoscale.

We thank Stephen Ducharme and Hermann Kohlstedt for stimulating discussions. This work was supported by NSF (grants DMR-0203359 and MRSEC: DMR-0213808), Nebraska Research Initiative, and Army Research Office. Computations were performed utilizing the Research Computing Facility of the University of Nebraska-Lincoln.

* e-mail:tsymbal@unl.edu